\documentclass{article}

\usepackage{microtype}
\usepackage{graphicx}
\usepackage{subfigure}
\usepackage{booktabs} 

\usepackage{amsmath,bm,amsfonts}
\usepackage{caption}
\usepackage{float}

\usepackage{hyperref}

\usepackage[accepted]{icml2020}

\icmltitlerunning{
Bayesian Networks for Brain-Computer Interfaces: A Survey
}

\begin{document}

\twocolumn[
\icmltitle{
Bayesian Networks for Brain-Computer Interfaces: A Survey
}




\begin{icmlauthorlist}
\icmlauthor{Pingsheng Li}{mcgill}
\end{icmlauthorlist}

\icmlaffiliation{mcgill}{School of Computer Science, McGill University, QC, Canada}

\icmlkeywords{Machine Learning, ICML}
\vskip 0.3in
]




\begin{abstract}
Brain-Computer Interface (BCI) is a rapidly developing technology that allows direct communications between the human brain and external devices, such as robotic arms and computers. Bayesian Networks is a powerful tool in machine learning for tackling with problems that requires understanding and modelling the uncertainty and complexity within complex system built by sub-modular components. Therefore, deploying Bayesian Networks in the application of Brain-Computer Interfaces becomes an increasingly popular approach in BCI research. This survey covers related existing works in relatively high-level perspectives, classifies the models and algorithms involved, and also summarizes the application of Bayesian Networks or its variants in the context of Brain-Computer Interfaces.
\end{abstract}

\section{Introduction}

In recent years, Brain-Computer Interface (BCI) has become a highly active research topic promoted by the rise of machine learning techniques. Due to its interdisciplinary nature, efficient BCI requires incorporating knowledge in Neuroscience with robust quantitative approaches in machine learning ~\cite{muller04}. Bayesian Networks, as a pivotal member in the probabilistic graphical model family, has been proven to be a very powerful tool in analyzing even very noisy neural data, such as EEG, MRI, and fMRI, etc.~\cite{bielza14}, which has greatly facilitated research in neuroscience. 

Therefore, it's natural to consider utilizing Bayesian Networks model or its variants, e.g. Dynamic Bayesian Networks, for decoding highly noisy neural signal and performing inference on the intention of users in Brain-Computer Interfacing, thereby improving the performance of BCI. There are already a few attempts on this approach ~\cite{NIPS2004_daaaf136},~\cite{10.3389/fnhum.2021.788258},~\cite{NIPS2017_99c5e07b},~\cite{DBLP:journals/corr/SalehiMNHE17},~\cite{7167726},~\cite{8342691}, ranging from cursor manipulation via motor imagination to channel selection for signal processing, etc. This survey attempts to cover relevant existing and more recent works of applying Bayesian Networks into BCI technology in high-level perspectives. Section 2 generally classifies the Bayesian Networks commonly used in BCI research. Section 3 describes inference and learning algorithms for the previously reviewed Bayesian Networks models applied in BCI. Section 4 summarizes the application of these Bayesian Network models and algorithms in BCI technology. Finally, section 5 concludes the survey by discussing the strength and limitation of Bayesian Networks or its variants, and hopefully offers promising directions for future research of the application of Bayesian Networks for BCI. 

\section{Models}

This section briefly introduces and summarizes a wide range of Bayesian Network models commonly used in BCI research.

\subsection*{Simple Bayesian Networks}
\begin{figure}[ht]
\centering
\includegraphics[width=0.6\linewidth]{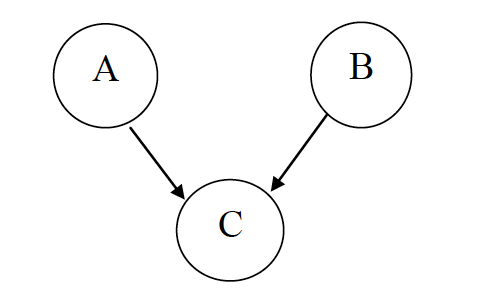}
\caption{A Simple Bayesian Network}
\label{simple}
\end{figure}

Bayesian Network (BN) is a probabilistic graphical model structure which consists of a node set and a directed edge set. The nodes represent random variables and the edges represent direct dependencies among these random variables. Notably, Bayesian Networks are required to be directed acyclic graphs (DAGs). See Fig.\ref{simple}.

Given a Simple Bayesian Network modelling a set of variables $\mathbf{X}=\{X_1, \dots, X_n\}$, we can directly read out the local conditional independence among variables from graphical structure, so the joint probability distribution can be factorized as the product of conditional probabilities of each variable $X_i$ given its parents $Pa_{X_i}$:
\begin{align*}
    P(\mathbf{X})=\prod_{i}P(X_i|Pa_{X_i})
\end{align*}
whereas without a Bayesian Network, by chain rule, the joint probability distribution will be written as:
\begin{align*}
    P(\mathbf{X})=P(X_1)P(X_2|X_1)\dots P(X_n|X_{n-1}, \dots, X_1)
\end{align*}
which significantly increases the number of parameters required for describing the exact same joint probability distribution.

\subsection*{Gaussian Bayesian Networks}
\begin{figure}[ht]
\centering
\includegraphics[width=0.6\linewidth]{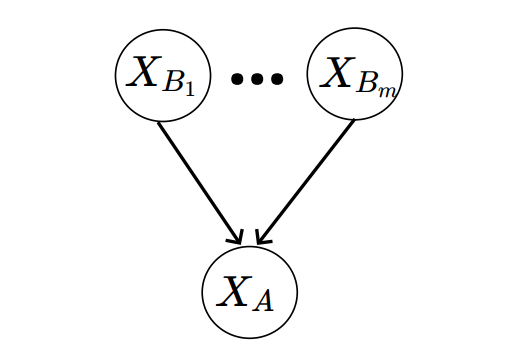}
\caption{A Gaussian Bayesian Network.}
\label{GBN}
\end{figure}

Gaussian Bayesian Networks (GBN) gives an alternative representation for multivariate Gaussian distribution in the form of Bayesian Networks. GBN requires all variables to be defined by a Gaussian prior distribution or a Gaussian conditional distribution, whose mean is a linear combination of mean of the parent variables with constant variance. See Fig.\ref{GBN}

For example, in Fig.\ref{GBN}, $\mathbf{X}_B \sim N(\mu_B, \Sigma_B)$, let $\mathbf{w}$ be a weight vector, then:
\begin{align*}
    X_A \sim N(\mathbf{w}^T\mu_B, \mathbf{w}^T\Sigma_B\mathbf{w})
\end{align*}
Another worth-noticing fact is that the resulting distribution of $X_A$ only has a single mode, in contrast to other mixture models which may have multiple modes in their distributions.

\subsection*{Gaussian Mixture Bayesian Networks}
Gaussian Mixture Model describes a distribution whose probability density function (PDF) is a linear combination of PDFs of different Gaussian distributions, which can also be represented as a similar Bayesian Network like GBN. But unlike GBN, a distribution characterized by Gaussian Mixture Model has multiple modes.

\subsection*{Dynamic Bayesian Networks}
\begin{figure}[ht]
\centering
\includegraphics[width=1\linewidth]{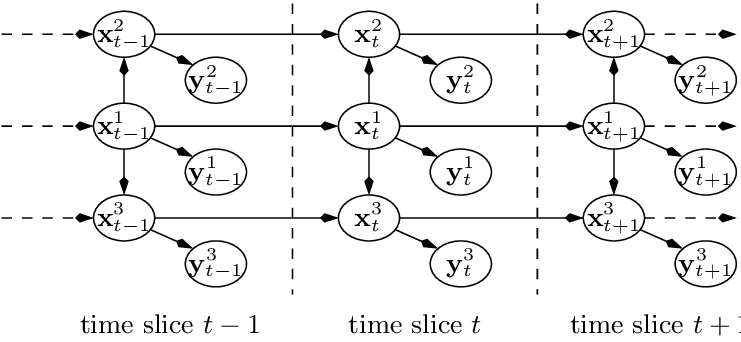}
\caption{A Dynamic Bayesian Network.}
\label{DBN}
\end{figure}

The Bayesian Network models discussed so far are static. So one problem of these models is that in systems evolving over time, such models will be unable to fully describe the interactions inside the system. To solve this problem, we need Dynamic Bayesian Network. Dynamic Bayesian Network (DBN) is a variant of Bayesian network designed for modeling data generated by systems with rich temporal features, such as time series, where all variables relates to each other (and often themselves) over consecutive time steps. A discrete time-stamp is introduced and the same local model, a section of the network called a \textit{time slice} that represents a snapshot of the underlying process, is repeated for each unit of time. It is of great convenience for applying Dynamic Bayesian Networks to analyze time series because there's an assumption of time series modeling that can greatly simplify the deign of Bayesian Networks: an event occurred at time $t$ can affect another event at time $t+1$ or later, but not vise-versa, so the directed edges should only flow forward in time. Therefore, one doesn't need to worry about breaking the acyclicity requirement for Bayesian Networks as long as the local model is acyclic. Fig.\ref{DBN} provides an illustration of DBN.

\subsection*{Hidden Markov Models}
\begin{figure}[ht]
\centering
\includegraphics[width=1\linewidth]{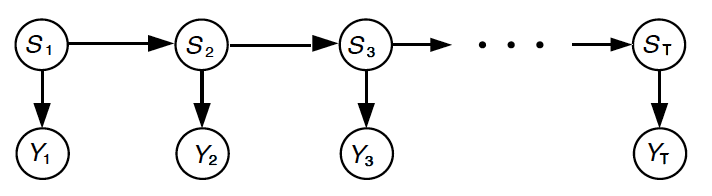}
\caption{A Dynamic Bayesian Network illustrating the conditional independence for a Hidden Markov Model.}
\label{HMM}
\end{figure}

A Hidden Markov Model (HMM) is designed for representing the probability distributions over sequences of observations, and it has two assumptions. First, each observation $Y_t$ at time $t$ is generated by some process whose state $S_t$ is hidden from the observer. Second, the hidden process satisfies the Markov property: given the current state $S_{t}$, the next state $S_{t+1}$ is independent of all prior states, i.e. $p(S_{t+1}|S_{t}, S_{t-1}, ...) = p(S_{t+1}|S_{t})$. In other words, the evolution of the process in the future depends only on the present state and does not depend on past history. HMM can also be considered as a special case of Dynamic Bayesian Network. See Fig.\ref{HMM}

\subsection*{Hidden semi-Markov Models}
A Hidden semi-Markov Model is a variant of HMM where the Markov property is relaxed to the semi-Markov property: The conditional distribution over the state at time $S_{t+1}$ depends not only on the current
state $S_{t}$, but also on a duration $d_t$ which encodes how long the state is to remain unchanged, i.e. $p(S_{t+1}|d_t, S_{t}, d_{t-1}, S_{t-1}, ...)=p(S_{t+1}|d_t, S_{t})$.

\subsection*{Common Bayesian Networks}
A Common Bayesian Network (CBN) can be described as a Bayesian Network built from common structural features from a collection of different Bayesian Networks. Sometimes, due to the massive noisiness in the data, the structures of the learned Bayesian Networks are highly unstable and often vary from time to time. In order to procure an unique and robust structure for a Common Bayesian Network from a collection of highly variable Bayesian Networks learned from noisy data, all edges and nodes in these BNs are evaluated by certain graph statistics, such as Common Edge Rate (CER) and Node Variation Rate (NVR), which will be utilized later for the construction of a Common Bayesian Network.

\section{Algorithms}
This section in general covers the algorithms applied in the models introduced above.

\subsection{Inference}
Inference in Bayesian Networks is to solve a probability when a Bayesian Network graphical structure is already given. Inference methods could be exact or approximate.

\subsubsection{Brutal Force Inference}
Brutal force inference approach is to directly solve the marginal or conditional probability required by summing over the probability mass functions or numerically integrating the probability density functions of a joint probability distribution, then applying Bayesian rule and marginalization: 
\begin{align*}
    p(x|e) = \frac{p(x, e)}{p(e)} \propto \sum_{u}p(x, e, u) \hspace{2mm} or \int_{u}p(x, e, u)
\end{align*}
where $x$ means the variables of interest, $e$ represents the observed evidence variables, and $u$ denotes other variables. This approach is algorithmically extremely inefficient, but sometimes acceptable when the requirement in computing time is not strict.

\subsubsection{Maximum A Posteriori (MAP)}
Maximum A Posteriori inference (MAP) is to solve which assignment of values variables $\theta$ will maximize a conditional probability $f(\theta|x)$, given that the evidence variables $x$ are observed, when the distribution is already known. In short, it aims to solve:
\begin{align*}
    \hat{\theta}_{MAP}(x)=\mathrm{argmax}_{\theta}f(\theta|x)
\end{align*}
Notably, both exact and approximate MAP inferences are NP-hard problems in general. However, many techniques, such as graph cuts and linear-programming, may sometimes offer efficient solutions. Also, there are special cases where MAP inference in Bayesian Networks can be performed very efficiently. For example, in hidden Markov models (HMMs), the most probable sequence of hidden variables can be computed using Viterbi algorithm, which performs a single pass of max-product inference over the model. \cite{NIPS2017_99c5e07b}.

\subsubsection{Junction Tree and Message Passing}
The junction tree algorithm, also known as Clique Tree, first partitions the Bayesian Network into clusters of variables where internally, the variables within a cluster could be highly coupled. Nevertheless, interactions among all clusters will have a tree structure, i.e., a cluster will be only directly influenced by its neighbors in the tree. This can give more tractable global solutions if the local, cluster-level problems can be solved exactly. And then, a message passing algorithm will be able to efficiently solve the probability required in the query, where some probabilities are initialized as messages and they will eventually converge to the true probabilities as these messages are passing among clusters. Therefore, it's an efficient approximate inference algorithm.

\subsection{Learning}
Learning in Bayesian Networks is to fit a model that will make predictions on various tasks relevant to the problem when a dataset is given, which includes both parameter learning, which is to estimate the parameters describing the distribution when the graph structure is known, and structure learning, which is to estimate the underlying directed acyclic graph , i.e., determine the variables dependencies from data.

\subsubsection{Maximum Likelihood Estimation (MLE)}
Maximum Likelihood Estimation is a parameter learning method whose goal is to find that for which values of model parameters $\theta$, the likelihood function $f(\mathbf{x}|\theta)$ over the parameter space will be maximized for data $x$ in dataset $D$. To summarize, it aims to exactly solve:
\begin{align*}
    \hat{\theta}_{MLE}=\mathrm{argmax}_{\theta}f(\mathbf{x}|\theta)
\end{align*}

\subsubsection{Expectation Maximization (EM)}
Expectation-Maximization algorithm is an iterative parameter learning algorithms for directed latent-variable graphical models $p(x, z; \theta)$ with observed data $x \in$ dataset $D$, parameters $\theta$, and latent $z$ (never observed), which alternately updates a posterior $p(z|x;\theta_{t})$ (E-step) and the parameters based on the following rule (M-step):
\begin{align*}
    \theta_{t+1}:=\mathrm{argmax}_{\theta}\sum_{x\in D}\mathrm{E}_{z\sim p(z|x;\theta_{t})}\log p(x,z;\theta)
\end{align*}
until convergence. 

\subsubsection{Score-based approach}
Score-based approach is for structure learning, which first defines a criterion to evaluate how well the Bayesian network fits the data, then searches over the space of DAGs for a structure achieving the maximal score, such as the Cheeseman–Stutz, Bayesian information criterion (BIC), and Laplace approximation scores, etc. 

\section{Applications}
This section covers the applications of the models mentioned above with more technical details.

\subsection*{Channel Selection}
The signal used for BCI system is collected via electrodes capturing brainwave activity, which is also called channels. However, not all signal channels are equally important in decoding neural data, and sometimes they can even confound the analysis. Therefore, it's extremely important to select the most effective channels for signal analysis. There should be two primary objectives in a BCI channel selection problem: (i) maximize the performance of the BCI system, and (ii) minimize the number of channels. Because these channels often are closely correlated, it is unnecessary to keep all of them, so some works choose to take advantage of simple Bayesian Networks to identify the most probable channels associated with certain tasks by modeling the dependencies among BCI channels and only using these relevant channels for later decoding ~\cite{7167726}, \cite{8342691}. In their works, they treat each EEG channel as a node of a simple Bayesian Network. The user could perform certain tasks, e.g.  wheelchair control, lamp control, and robotic arm manipulation, etc., by just imagining the corresponding movements (motor imagery, MI). During the imagination of a movement there will be correlation between several EEG channels. For a particular movement, certain channels may be dependent whereas others may be independent. These dependence and independence can be modeled as the nodes and edges of a Bayesian Network. Using this network, the most probable channels for each movement could be estimated. The exact procedure is described below.
\subsubsection*{1. GMM Construction for Channels with EM}
First, use Gauassian Mixture Model (GMM) to parameterize the data points collected by each channel. For an EEG channel signals with a set of $N$ points in $D$ dimensions, $x_1,\dots, x_N \in R^D$ and all mixtures of Gaussian functions $P$, the task is to find the probability density function $p(x) \in P$ that is most likely to generate the given data points. Functions in $P$ can written as:
\begin{align*}
    p(x;\theta) &= \sum^{K}_{l=1} a_k g(x; m_k, \delta_k) \\
    g(x; m_k, \delta_k) &= \frac{1}{\sqrt{2\pi}\delta_k}e^{-\frac{1}{2}(\frac{||x-m_k||}{\delta_k})^2}
\end{align*}
where $x$ is a $D$-dimensional continuous-valued data vector,
$\theta = (\theta_1,\dots,\theta_k)=((a_1,m_1,\delta_1),\dots,(a_k,m_k,\delta_k))$ is a $K(D + 2)$-dimensional vector, where $a_k, k \in \{1, 2,\dots,K\}$ is the weights of $k^{th}$ mixture model satisfying $\sum^{K}_{i=1}a_{i} = 1$. $g(x;m_k, \delta_k)$ is a $D$-dimensional isotropic Gaussian function with mean vector $m_k$ and the standard deviation $\delta_k$. In order to calculate $\theta$ that gives the highest probability, the likelihood function is defined as:
\begin{align*}
    L(X; \theta)=\prod^{N}_{n=1}f(x_n;\theta)=\prod^{N}_{n=1}\sum^{K}_{l=1} a_k g(x; m_k, \delta_k)
\end{align*}
So the task becomes to find the best parameters $\hat{\theta}$:
\begin{align*}
    \hat{\theta} = \mathrm{argmax}_{\theta}L(X; \theta)
\end{align*}
Then the problem can be solved using expectation maximization algorithm (EM). First, initialize values $a^{0}_k, m^{0}_k, \delta^{0}_k$, then iteratively repeat E step and M step below until convergence as described in section 2.\\
\textit{E step:}
\begin{align*}
    a^{(i)}(k|n)=\frac{a^{(i)}_{k}g(x_n; m^{(i)}_{k}, \delta^{(i)}_{k})}{\sum^{K}_{k=1}a^{(i)}_{k}g(x_n; m^{(i)}_k, \delta^{(i)}_k)}
\end{align*}
\textit{M step:}
\begin{align*}
    m^{(i+1)}_k &= \frac{\sum^{N}_{n=1}a^{(i)}(k|n)x_n}{\sum^{N}_{n=1}a^{(i)}(k|n)}\\
    \delta^{(i+1)}_k &= \sqrt{\frac{1}{D}\frac{\sum^{N}_{n=1}a^{(i)}(k|n)||x_n - m^{(i+1)}_k||^2}{\sum^{N}_{n=1}a^{(i)}(k|n)}}\\
    a^{(i+1)}_{k}&=\frac{1}{N}\sum^{N}_{n=1}a^{(i)}(k|n)
\end{align*}
Exact derivation is omitted. After this step, the GMM model for the signal of each channel is constructed.
\subsubsection*{2. Conditional Densities for Two GMMs}
After constructing a GMM model for each channel, one can find their conditional probability via Bayes' rule:
\begin{align*}
    p(y|x) = \frac{p(x, y)}{p(x)}
\end{align*}
For simplifying the calculations, one may need a few assumptions \cite{7167726}): the number of mixture for GMMs should be limited (e.g. $K=2$); the covariance matrix should be constrained to be diagonal; the components of Gaussian mixtures are independent. Given these assumptions, because here both $x$ and $y$ are mixtures of Gaussian distributions, therefore:
\begin{align*}
    p(y,x) &= \sum^{n}_{i=1}a_i N(y, m_{y,i}, \delta_{y,i}) N(x, m_{x,i}, \delta_{x,i})\\
    p(y|x) &= \sum^{n}_{i=1}W_i(x) N(y, m_{y,i}, \delta_{y,i})
\end{align*}
where 
\begin{align*}
    W_i(x) &= \frac{a_i N(x, m_{x,i}, \delta_{x,i})}{\sum^{n}_{j=1}a_j N(y, m_{y,j}, \delta_{y,j})}
\end{align*}
\subsubsection*{3. Greedy Search from Scores Based on GMM Nodes}
The last step of construction a Bayesian Network for BCI channels is to find the structure of the network that gives the best score used for evaluating the quality of Bayesian Networks, such as Cheeseman-Stutz Criterion, Bayesian Information Criterion (BIC), and Laplace approximation scores. However, the most commonly used score is BIC, because it not only measures the efficiency of the model, but also penalizes the complexity (the number of parameters) of the model, therefore prevents learning very densely connected networks that are meaningless for the original intention of improving channel efficiency \cite{7167726}. The BIC score can be calculated according to the likelihood function:
\begin{align*}
    BIC= -2 \hspace{0.5mm}ln \hspace{0.5mm} p(x|M) + L \hspace{0.5mm} (ln \hspace{0.5mm} n - ln \hspace{0.5mm} 2\pi)
\end{align*}
where $L$ is the number of free parameters to be estimated. Because there are three parameters for each Gaussian component, i.e., $a, \delta, m$, so $L = 3K$ where K is the number of Gaussian components. $n$ is the number of data points in $x$, and $p(x|M)$ is the marginal likelihood of the observed data given model M with certain structure (some edges between nodes). Finally, the structure with the minimum BIC score constructed based on the greedy search algorithm is chosen as the final result of learning. \\

\subsubsection*{4. (Optional) Build Common Bayesian Networks}
Although after step 3 one could already obtain a collection of valid Bayesian Networks, sometimes these learned networks can still have very unstable structures due to the noisiness of the neural data. In order to solve this problem, \citet{7167726} designed an algorithm that can build Common Bayesian Networks that characterize the common features among a set of different Bayesian Networks. Define:
\begin{align*}
    E_k(i, j) &= \begin{cases}
  0  & \text{if} (i, j) \notin E_k \\
  1  & \text{if} (i, j) \in E_k \\
\end{cases} \\
    Cr(i, j) &= \sum^{N}_{k=1}\frac{E_k (i,j) + E_k (j,i)}{2N}
\end{align*}
where $N$ is the total number of Bayesian Networks obtained before, and $E_k$ is the edge set of BN $G_k$. $Cr(i, j)$, called edge common rate, can be viewed as the probability of an edge from $i$ to $j$. Also, for $i^{\text{th}}$ run of experiments and $j^{\text{th}}$ node, define:
\begin{align*}
    W_i &= (w_{i1}, \dots, w_{iN})\\
    w_{ij} \hspace{1mm} +\hspace{-1mm}&= \begin{cases}
  0  & \text{if no edge on node } j \text{ in } G_i \\
  1  & \text{if node } j \text{ is a parent in } G_i \\
  -1  & \text{if node } j \text{ is a child in } G_i \\
\end{cases} \\
    m_i &= \frac{1}{N_i}\sum_{w \in W_i}w, i=1, 2, \dots, Q \\
    \alpha_i &= \sum_{w \in W_i}(w-m_i)(w-m_i)^T, i=1, 2, \dots, Q \\
    \beta_{ij} &= (m_i - m_j)(m_i - m_j)^T \\
    f &= \frac{\sum^{N_i}_{i=1}\alpha_i}{\sum^{N_i}_{i=1}\sum^{P}_{j=1}\beta_{ij}}
\end{align*}
where $N_i$ is the run count of the $i^{\text{th}}$ MI task and $Q$ is the number of MI tasks. $\alpha$ is the intraclass distance matrix, and $\beta_{ij}$ is the interclass distance matrix for two MI task $i$ and $j$. Every node has P intraclass distances and P×P interclass distances. $f$ is the node variation rate described by the ratio of $\alpha$ and $\beta$. The larger it is, the more sensitive it is to MI. In other words, if one node has a large $f$ , this node has a large difference between interclass and intraclass of MI tasks, so this node is a key node since it's stable and discriminative. In the final Common Bayesian Network, one can only choose edges with edge common rate $Cr(i,j)>\delta_0$ and nodes with node variation rate $f>f_0$ as the skeleton of the network, where $\delta_0$ and $f_0$ are certain manually determined threshold values. Other structural constraints on the learned Bayesian Networks, such as topological adjacency, etc., may also be applied.

\begin{figure}
    \centering
    \includegraphics[width=1\linewidth]{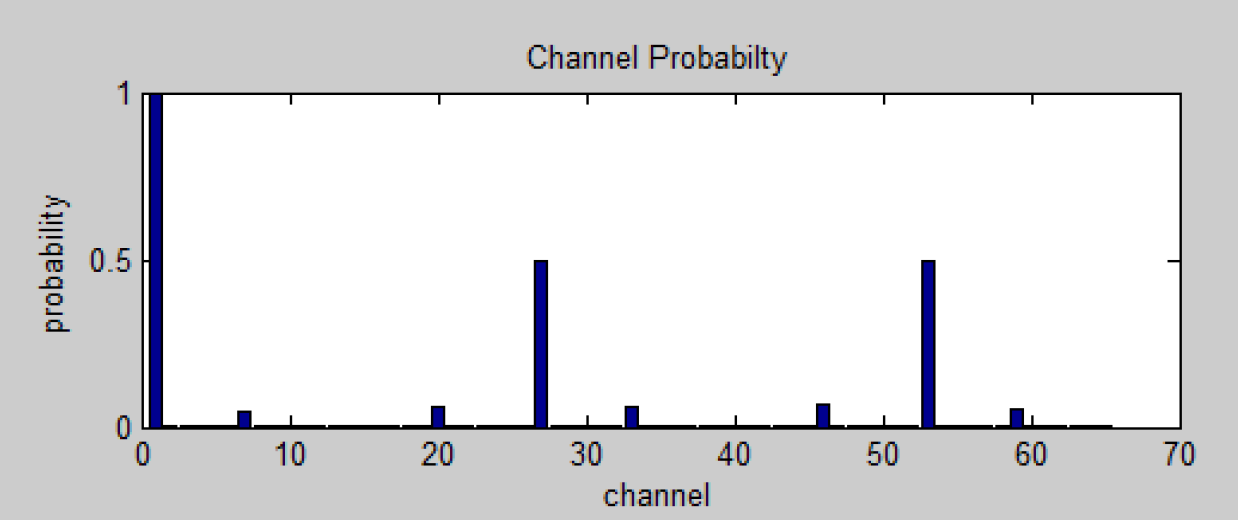}
    \caption{EEG channel probability during the left fist movement. The EEG system used in the experiment has 64 channels, and only 3 those channels with probability higher than certain threshold are used for decoding certain movements. Reprinted from \citet{8342691}}
    \label{fig:CS}
\end{figure}
\newpage
Applying Bayesian Networks for channel selection is highly efficient. It can significantly reduce the number of channels required for each task, and in turn drastically improve computational efficiency and performance of BCI system. \citet{8342691} showed that by utilizing Bayesian Networks, even in EEG system with 64 channels, only 3 channels are needed for decoding certain movements, which greatly reduces the computing power required and allows more efficient, real-time decoding (See Fig.\ref{fig:CS}).

\subsection*{Classification}
\begin{figure}[h]
    \centering
    \includegraphics[width=\linewidth]{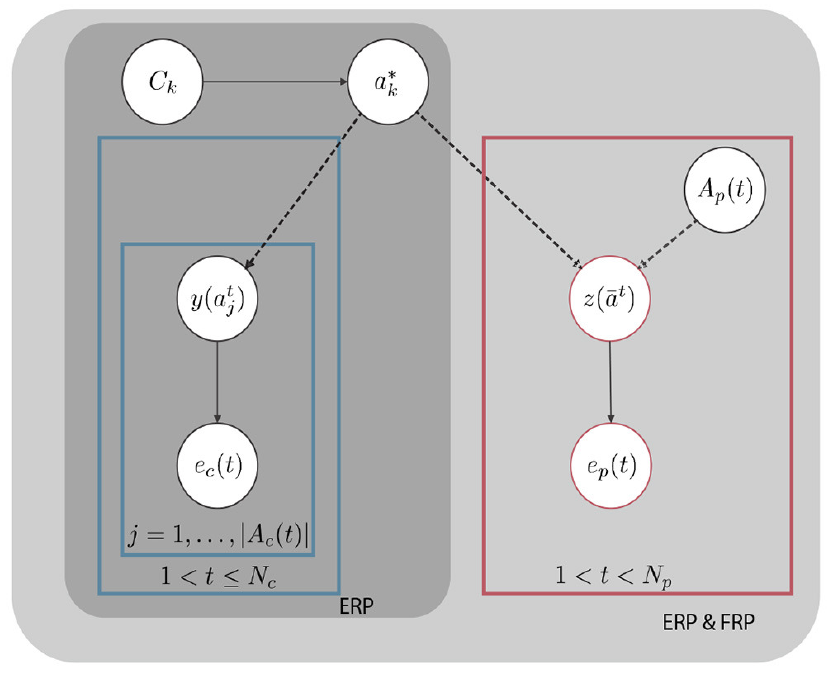}
    \caption{Proposed probabilistic graphical model representing the $k^{th}$ epoch of data. Here, the dashed lines show a deterministic relation while the solid lines define a probabilistic correspondence. $z ( \Bar{a}^t ) = 1$ ErrP label, $z ( \Bar{a}^t ) = 0$ non-ErrP label. $y (a^t_j ) = 1$ target label, $y (a^t_j ) = 0$ non-target label. $t$ denotes sequence index. $j$ denotes trial index. Reprinted from  \cite{10.3389/fnhum.2021.788258}}
    \label{fig:ERP}
\end{figure}

\begin{figure}
    \centering
    \includegraphics[width=1\linewidth]{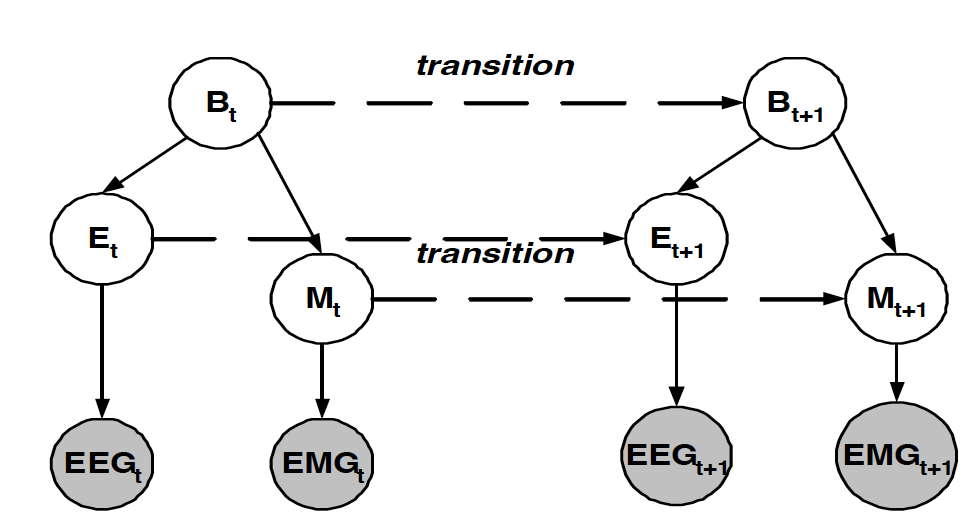}
    \caption{Dynamic graphical model for modeling brain and body processes in a self-paced movement task: (a) At each time instant $t$, the brain state $B_t$ generates the EEG and EMG internal states $E_t$ and $M_t$ respectively, which in turn generate the observed EEG and EMG. The dotted arrows represent transitions to a state at the next time step. Reprinted from \cite{NIPS2004_daaaf136}}
    \label{fig:DBN2}
\end{figure}

The BCI system needs to decode user's intents from a set of actions in various aspects, e.g. motor imagery, visually evoked potential, etc., which is essentially a classification problem. Also, given a collection of recorded signal, such as EEG, EOG, the classification task can effectively be converted into an inference problem: decide which is the most probable brain state (intent) that corresponds to certain action(s) given the observed signal. Such decision making process requires performing maximum a posteriori (MAP) inference on a given Bayesian Network structure. A common design of such a Bayesian Network is to set the brain state (intent) node as the parent of the nodes of other variables that generates the recorded signals using a mixture of Gaussian distributions conditioned on them. Similarly, conditioned on the brain state nodes, the distribution of these generating nodes can also be modelled via Gaussian Mixture or Naive Bayes Model. Fig.\ref{fig:ERP} and Fig.\ref{fig:DBN2} provide illustrations of such design in the context of Dynamic Bayesian Networks.

After the parameters are learned via expectation maximization (EM) or other kernel estimation methods, the MAP inference can be performed using various algorithms depending on the network structure. For example, one can utilize the junction tree algorithm on slightly complicated Dynamic Bayesian Networks to infer mental states in order to send command to the external devices \cite{NIPS2017_99c5e07b}, \cite{Saa2014ProbabilisticGM}, \cite{DBLP:journals/corr/abs-1809-05635}. The complexity of junction tree algorithm depends on the tree-width, a measure of similarity between the graph and a tree, of the network, which makes inference on slightly complex networks still tractable despite the fact that both exact and approximate MAP inference problems are NP-hard.

\begin{figure}[H]
    \centering
    \includegraphics[width=0.75\linewidth]{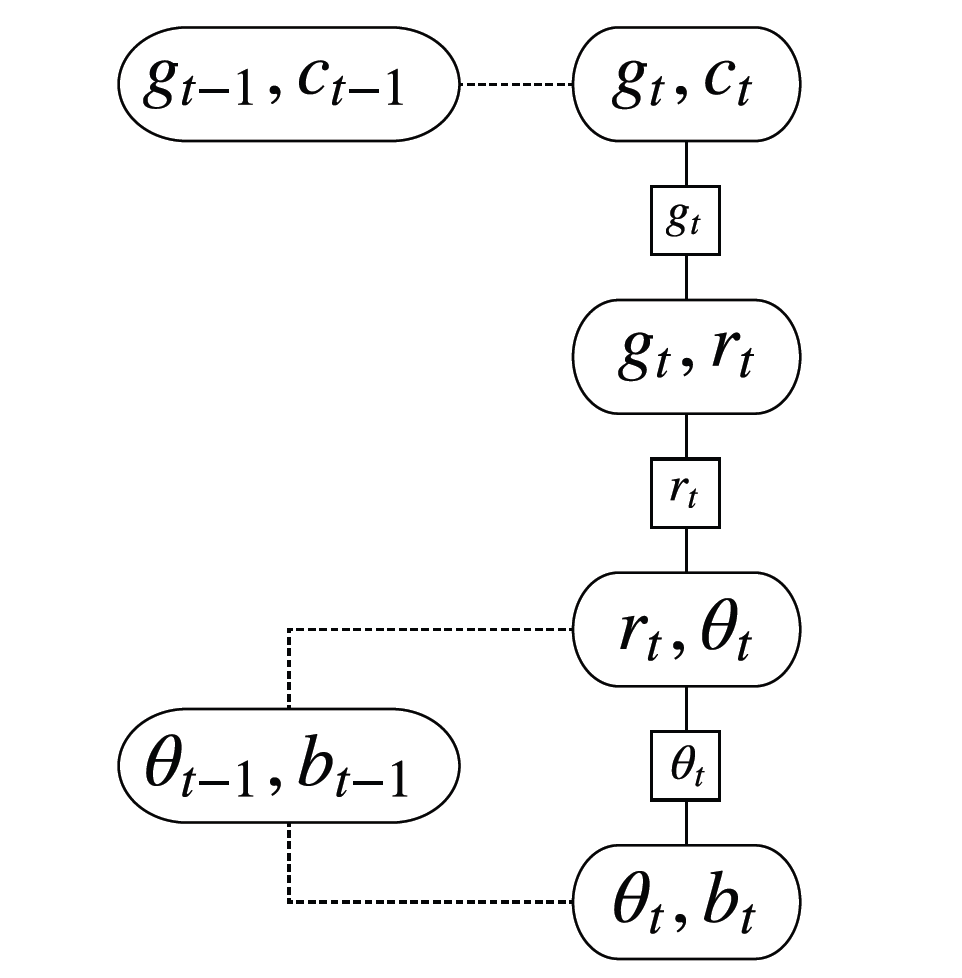}
    \caption{Illustration of the junction tree used to compute marginals for online decoding. Dashed edges indicate cliques whose potentials depend on the marginal approximations at time $t-1$. The inference uses an auxiliary variable $r_t ;= a(g_t; p_t)$ to reduce computation and allow inference to operate in real time. Reprinted from \cite{NIPS2017_99c5e07b}}
    \label{fig:JT}
\end{figure}

\vspace{-3mm}
For Hidden Markov Models or networks with similar structures (Fig.\ref{fig:DBN2}), one can perform the MAP inference even faster using Viterbi algorithm (a single pass of max-product inference over the network) \cite{NIPS2004_daaaf136}. Sometimes, even local search method for MAP is also acceptable.

However, there is no existing work that uses linear programming approach or sampling methods, such as simulated annealing, Metropolis-Hastings, etc., for MAP inference in the context of BCI system, which could be a possible direction for future research.

\subsection*{Dynamics Modeling}
\begin{figure}
    \centering
    \includegraphics[width=1\linewidth]{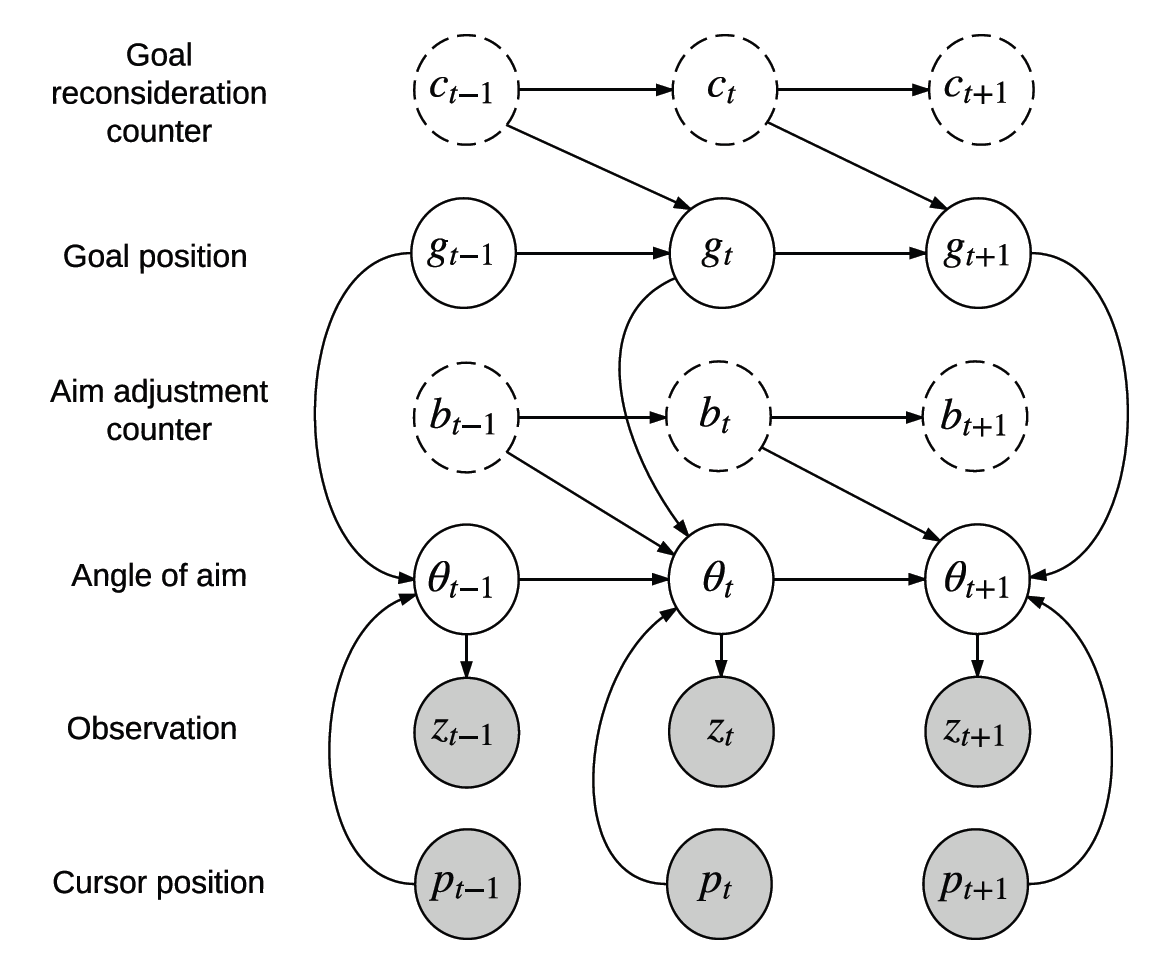}
    \caption{The multi-scale directed graphical model for a cursor controlling BCI illustrating how goal positions $g_t$, angles of aim $\theta_t$, and observed cursor positions $p_t$ evolve over three time steps. Dashed nodes are counter variables enabling semi-Markov dynamics. Reprinted from \cite{NIPS2017_99c5e07b}}
    \label{fig:semi}
\end{figure}

Bayesian Networks, specifically Dynamic Bayesian Networks, are also very powerful for modeling systems with temporal dynamics. One reason is that it's naturally compatible with the Markov property, i.e. $p(S_{t+1}|S_{t}, S_{t-1}, ...) = p(S_{t+1}|S_{t})$. Hidden Markov Model is the example that perfectly incorporate the Markov property into Bayesian Networks (See Fig.\ref{HMM}). More importantly, one could easily extend this property to $k^{\text{th}}$ order Markov property:
\begin{align*}
    p(S_{t+1}|S_{t}, S_{t-1}, ...) = p(S_{t+1}|S_{t}, S_{t-1}, \dots, S_{t-k+1})
\end{align*}
by simply connecting more nodes $S_{t-1}, \dots, S_{t-k+1}$ to $S_{t}$. Furthermore, one could also use Dynamic Bayesian Networks when the Markov property is relaxed to semi-Markov property:
\begin{align*}
   p(S_{t+1}|d_t, S_{t}, d_{t-1}, S_{t-1}, ...)=p(S_{t+1}|d_t, S_{t})
\end{align*}
where $d_t$ is the time spent on state $S_t$, by just adding one more variable $d_t$ at each time slice $t$ and its corresponding edges into the networks. Similar to incorporating $k^{th}$ order Markov property into the networks, DBNs with higher semi-Markov property can also be constructed similarly. See Fig.\ref{fig:semi} for an example.

Both Markov \& semi-Markov properties are very important assumptions when people are dealing with sequential decision problems in a dynamical system. BCI also often needs to solve such sequential decision problems in real time while keeps interacting with the environment. For example, a BCI may need to send a sequence of commands to control a wheelchair to proceed, stop, and turn around within a few seconds, which could be viewed as a Markov Decision Process (MDP). In tasks that requires even more delicate controls, such as moving the cursor of a computer, the Markov property must be relaxed to semi-Markov property in order to achieve more desirable performance \cite{NIPS2017_99c5e07b}.

Dynamic Bayesian Networks, therefore, can successfully tackle with such problems because: (i) The posterior probability could be factorized or partially factorized by utilizing the conditional independence encoded by the network structure, making MAP-inference-based decision making more tractable and computationally efficient. (ii) They can be easily extended to incorporate higher order Markov \& semi-Markov properties, which is very flexible when dealing with more complex dynamical environment.

One example of such application is cursor controlling intracortical BCI for patients with paralyzed arms or hands \cite{NIPS2017_99c5e07b}. Before this work, similar BCIs are based on a Kalman filter that assumes the vector of desired cursor movement evolves according to Gaussian random walk dynamics, and that neural activity is a Gaussian-corrupted linear function of this state. 
\newpage
\begin{figure}
    \centering
    \includegraphics[width=1\linewidth]{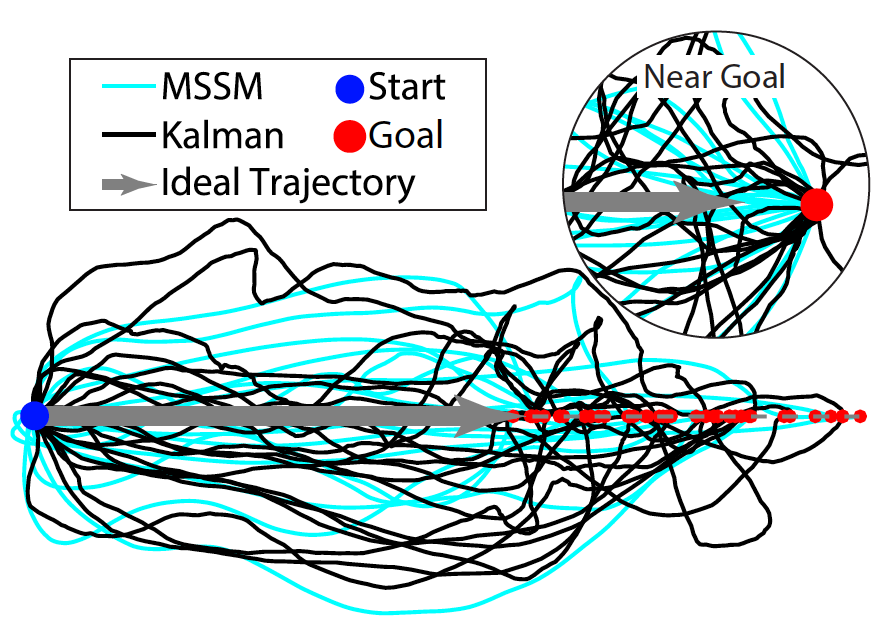}
    \caption{Examples of real-time decoding trajectories. 20 randomly selected trajectories for the Kalman decoder, and 20 trajectories for the MSSM decoder. The trajectories are aligned so that the starting position is at the origin and rotated so the goal position is on the positive, horizontal axis. The MSSM decoder exhibits fewer abrupt direction changes. Reprinted from \cite{NIPS2017_99c5e07b}}
    \label{fig:cursor}
\end{figure}

\begin{figure}
    \centering
    \includegraphics[width=1\linewidth]{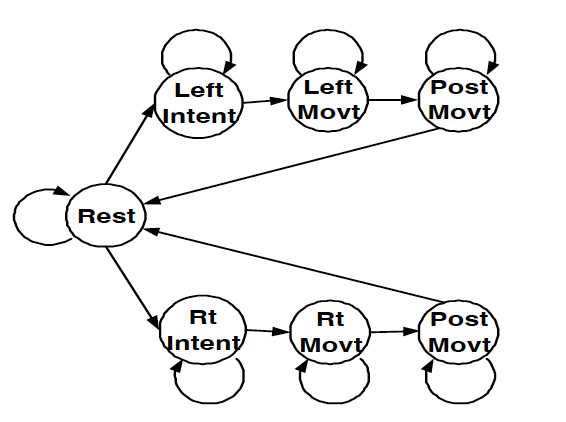}
    \caption{The transition graph for the brain state $B_t$. The probability of each allowed transition is learned from input data. Reprinted from  \cite{NIPS2004_daaaf136}}
    \label{fig:transition}
\end{figure}

However, this weak temporal dependence given by the first-order Markov assumption is highly problematic in the BCI setting due to \textbf{the mismatch between down-sampled sensor acquisition rates used for decoding} (typically around 50Hz, or 20ms per timestep) and the time scale at which the desired goal position changes (seconds). Therefore, relaxing the Markov property to semi-Markov property greatly improve the performance of their decoder called MSSM (See Fig.\ref{fig:cursor}). Also, one could customize the type of transitions that are allowed in the model to incorporate additional constraints that may help the performance \cite{NIPS2004_daaaf136}. See Fig.\ref{fig:transition} for illustration.

Therefore, it has been demonstrated that Dynamic Bayesian Networks are able to successfully model system dynamics that is critical to the sequential decisions making process in BCI.

\subsection*{Augmentation by Language Modeling}
\begin{figure}
    \centering
    \includegraphics[width=1\linewidth]{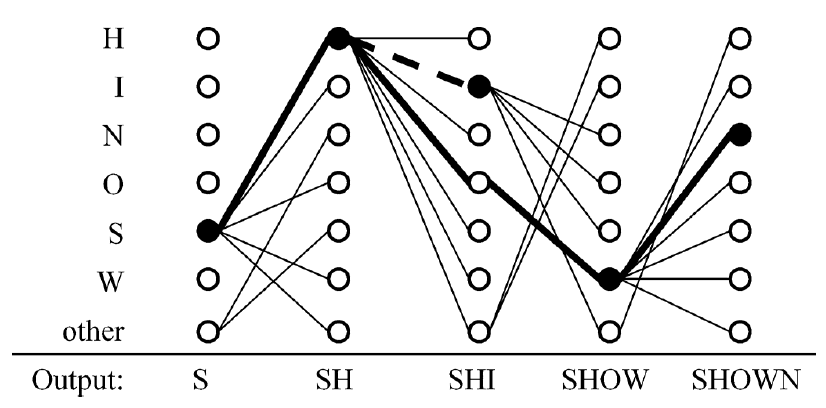}
    \caption{Simplified Viterbi trellis for subject B spelling the word "shown." At time $t=3$, the character "I" has the highest probability, resulting in the output "shi" after following the back pointers (dotted lines). At time $t=4$, the character "W" has the highest probability and the back pointers (bold lines) produce the output "show," correcting the previous mistake. Reprinted from \cite{Speier2014IntegratingLI}}
    \label{fig:lang}
\end{figure}

A huge portion of non-invasive, EEG-based BCI aims to facilitate language communication, including letter-by-letter typing BCI system. Famous examples are P300 spellers, steady-state visual evoked potential (SSVEP) spellers, etc. However, since the signal acquired by non-invasive method is often highly noisy, several trials must be combined in order to correctly classify responses into letters. The resulting typing speed can therefore be very slow, prompting many reports focusing on system optimization. But before \citet{article} most attempts at system optimization have not taken advantage of existing knowledge about the language domain. Existing methods treat character selection problems as choosing some elements independently from a set with no prior information. However, in practice, information about the domain of natural language can be utilized to build a prior belief about the characters to be chosen, which could improve both speed and accuracy of the typing system. \citet{Speier2014IntegratingLI} applied Hidden Markov Models in language modeling to predict the next character and auto-correct previous incorrect characters in order to augment the performance of BCI spellers. They treat BCI communication as a hidden Markov model (HMM) where hidden states are target characters and the EEG signal is the visible output. Then using the Viterbi algorithm for MAP inference, language information can be effectively incorporated in classification and errors can be corrected automatically (See Fig.\ref{fig:lang}). 

Similarly, \citet{10.3389/fnhum.2021.788258} use Dynamic Bayesian Networks to build a n-gram language model for error related potential (ERP) based BCI spellers to incorporating context-related information (previous letters) in similar fashion when performing MAP inference to predict the current letter. Both has been shown that letter-level language modeling using Bayesian Networks can be highly successful in improving the efficiency and accuracy of common typing systems, such as EEG-based and ERP-based spellers, etc.

\subsection*{Multi-modal Integration}
Bayesian Networks are naturally suitable for modelling systems with cooperating sub-modular components, which can also facilitate the integration of information used for BCI from multiple modalities, such as EEG and EMG signals, and will allow, for example, EEG-derived estimates to be bootstrapped from EMG-derived estimates, which combines all information or signals from all modalities to achieve better performance \cite{NIPS2004_daaaf136}, \cite{NIPS2017_99c5e07b}.

Also, due to the noisiness nature of the neural data collected by non-invasive recording system, such as EEG, EMG, the signal could be unfaithful or even missing sometimes. In such cases, A dynamic Bayesian Network model for time-varying data such as EEG also allows prediction, filling in of missing data, and smoothing of state estimates using information from future data points. These properties cannot be easily achieved in methods that work exclusively in the frequency domain or use data slices for training classifiers. Therefore, Bayesian Networks have special superiority over other models in such applications. 

\section{Conclusion}
In conclusion, this survey summarizes relevant Bayesian Network models, algorithms, and their common applications for Brain-Computer Interface technology in high-level. The survey describes several very useful applications of Bayesian Networks and its variants in the development of BCI, such as channel selections for reducing computation and improving efficiency, MAP-inference based classification for decoding user's intents from recorded data, system-level dynamics modeling with Dynamic Bayesian Networks via Markov \& semi-Markov properties, language modelling for performance augmentation, and multi-modal integration of sub-modular components. Also, this survey finds that maximum-a-posteriori (MAP) inference is massively used in the application of Bayesian Networks for BCI. Therefore, developing more efficient MAP inference algorithm for BCI could be a promising direction for future research, and will greatly improve the performance of BCI system by allowing faster and more efficient decoding of the neural data. Linear programming approaches and sampling-based methods for MAP may offer interesting progress on this problem.

\bibliography{main}

\begin{thebibliography}{12}
\providecommand{\natexlab}[1]{#1}
\providecommand{\url}[1]{\texttt{#1}}
\expandafter\ifx\csname urlstyle\endcsname\relax
  \providecommand{\doi}[1]{doi: #1}\else
  \providecommand{\doi}{doi: \begingroup \urlstyle{rm}\Url}\fi

\bibitem[Bielza \& Larrañaga(2014)Bielza and Larrañaga]{bielza14}
Bielza, C. and Larrañaga, P.
\newblock Bayesian networks in neuroscience: a survey.
\newblock \emph{Frontiers in Computational Neuroscience}, 8, 2014.
\newblock ISSN 1662-5188.
\newblock \doi{10.3389/fncom.2014.00131}.
\newblock URL
  \url{https://www.frontiersin.org/article/10.3389/fncom.2014.00131}.

\bibitem[Gonzalez-Navarro et~al.(2022)Gonzalez-Navarro, Celik, Moghadamfalahi,
  Akcakaya, Fried-Oken, and Erdoğmuş]{10.3389/fnhum.2021.788258}
Gonzalez-Navarro, P., Celik, B., Moghadamfalahi, M., Akcakaya, M., Fried-Oken,
  M., and Erdoğmuş, D.
\newblock Feedback related potentials for eeg-based typing systems.
\newblock \emph{Frontiers in Human Neuroscience}, 15, 2022.
\newblock ISSN 1662-5161.
\newblock \doi{10.3389/fnhum.2021.788258}.
\newblock URL
  \url{https://www.frontiersin.org/article/10.3389/fnhum.2021.788258}.

\bibitem[He et~al.(2016)He, Hu, Wan, Wen, von Deneen, and Zhou]{7167726}
He, L., Hu, D., Wan, M., Wen, Y., von Deneen, K.~M., and Zhou, M.
\newblock Common bayesian network for classification of eeg-based multiclass
  motor imagery bci.
\newblock \emph{IEEE Transactions on Systems, Man, and Cybernetics: Systems},
  46\penalty0 (6):\penalty0 843--854, 2016.
\newblock \doi{10.1109/TSMC.2015.2450680}.

\bibitem[Milstein et~al.(2017)Milstein, Pacheco, Hochberg, Simeral,
  Jarosiewicz, and Sudderth]{NIPS2017_99c5e07b}
Milstein, D., Pacheco, J., Hochberg, L., Simeral, J.~D., Jarosiewicz, B., and
  Sudderth, E.
\newblock Multiscale semi-markov dynamics for intracortical brain-computer
  interfaces.
\newblock In Guyon, I., Luxburg, U.~V., Bengio, S., Wallach, H., Fergus, R.,
  Vishwanathan, S., and Garnett, R. (eds.), \emph{Advances in Neural
  Information Processing Systems}, volume~30. Curran Associates, Inc., 2017.
\newblock URL
  \url{https://proceedings.neurips.cc/paper/2017/file/99c5e07b4d5de9d18c350cdf64c5aa3d-Paper.pdf}.

\bibitem[Müller et~al.(2004)Müller, Krauledat, Dornhege, Curio, and
  Blankertz]{muller04}
Müller, K.-R., Krauledat, M., Dornhege, G., Curio, G., and Blankertz, B.
\newblock Machine learning techniques for brain-computer interfaces.
\newblock \emph{Biomed Tech}, 49, 01 2004.
\newblock \doi{10.13109/9783666351419.11}.

\bibitem[{\"{O}}zdenizci et~al.(2018){\"{O}}zdenizci, G{\"{u}}nay, Quivira, and
  Erdogmus]{DBLP:journals/corr/abs-1809-05635}
{\"{O}}zdenizci, O., G{\"{u}}nay, S.~Y., Quivira, F., and Erdogmus, D.
\newblock Hierarchical graphical models for context-aware hybrid brain-machine
  interfaces.
\newblock \emph{CoRR}, abs/1809.05635, 2018.
\newblock URL \url{http://arxiv.org/abs/1809.05635}.

\bibitem[Saa \& Fernando(2014)Saa and Fernando]{Saa2014ProbabilisticGM}
Saa, D. and Fernando, J.
\newblock Probabilistic graphical models for brain computer interfaces.
\newblock 2014.

\bibitem[Sagee \& Hema(2017)Sagee and Hema]{8342691}
Sagee, G.~S. and Hema, S.
\newblock Eeg feature extraction and classification in multiclass multiuser
  motor imagery brain computer interface u sing bayesian network and ann.
\newblock In \emph{2017 International Conference on Intelligent Computing,
  Instrumentation and Control Technologies (ICICICT)}, pp.\  938--943, 2017.
\newblock \doi{10.1109/ICICICT1.2017.8342691}.

\bibitem[Salehi et~al.(2017)Salehi, Moghadamfalahi, Nezamfar, Haghighi, and
  Erdogmus]{DBLP:journals/corr/SalehiMNHE17}
Salehi, S. S.~M., Moghadamfalahi, M., Nezamfar, H., Haghighi, M., and Erdogmus,
  D.
\newblock Context-aware recursive bayesian graph traversal in bcis.
\newblock \emph{CoRR}, abs/1703.02938, 2017.
\newblock URL \url{http://arxiv.org/abs/1703.02938}.

\bibitem[Shenoy \& Rao(2005)Shenoy and Rao]{NIPS2004_daaaf136}
Shenoy, P. and Rao, R.~P.
\newblock Dynamic bayesian networks for brain-computer interfaces.
\newblock In Saul, L., Weiss, Y., and Bottou, L. (eds.), \emph{Advances in
  Neural Information Processing Systems}, volume~17. MIT Press, 2005.
\newblock URL
  \url{https://proceedings.neurips.cc/paper/2004/file/daaaf13651380465fc284db6940d8478-Paper.pdf}.

\bibitem[Speier et~al.(2011)Speier, Arnold, Lu, Taira, and Pouratian]{article}
Speier, W., Arnold, C., Lu, J., Taira, R., and Pouratian, N.
\newblock Natural language processing with dynamic classification improves p300
  speller accuracy and bit rate.
\newblock \emph{Journal of neural engineering}, 9:\penalty0 016004, 12 2011.
\newblock \doi{10.1088/1741-2560/9/1/016004}.

\bibitem[Speier et~al.(2014)Speier, Arnold, Lu, Deshpande, and
  Pouratian]{Speier2014IntegratingLI}
Speier, W., Arnold, C.~W., Lu, J., Deshpande, A., and Pouratian, N.
\newblock Integrating language information with a hidden markov model to
  improve communication rate in the p300 speller.
\newblock \emph{IEEE Transactions on Neural Systems and Rehabilitation
  Engineering}, 22:\penalty0 678--684, 2014.

\end{thebibliography}
\bibliographystyle{icml2020}

\end{document}